\documentclass{article}
\usepackage{ijcai19}

\usepackage{times}
\usepackage{soul}
\usepackage{url}
\usepackage[utf8]{inputenc}
\usepackage[small]{caption}
\usepackage{graphicx}
\usepackage{amsmath}
\usepackage{booktabs}
\usepackage{subfigure}
\usepackage[ruled,linesnumbered]{algorithm2e} 
\usepackage{bm}
\usepackage{amsmath}
\usepackage{amssymb}
\usepackage{amsmath}

\newtheorem{myDef}{Definition} 
\newtheorem{lemma}{Lemma}

\usepackage{enumerate}
\usepackage{multirow}
\usepackage{booktabs}
\usepackage{float}
\title{Binarized Collaborative Filtering with Distilling Graph Convolutional Networks}

\author{
Haoyu Wang$^{1,2}$
\and
Defu Lian$^1$\footnote{Corresponding author}\and
Yong Ge$^{3}$
\affiliations
$^1$School of Computer Science and Technology, University of Science and Technology of China\\
$^2$University of Electronic Science and Technology of China\\
$^3$University of Arizona
\emails
\{dove.ustc, haoyu.uestc\}@gmail.com,
yongge@email.arizona.edu
}

\begin{document}
	
	\maketitle
	
	\begin{abstract}
		The efficiency of top-K item recommendation based on implicit feedback are vital to  recommender systems in real world, but it is very challenging due to the lack of negative samples and the large number of candidate items. To address the challenges, we firstly introduce an improved Graph Convolutional Network~(GCN) model with high-order feature interaction considered. Then we distill the ranking information derived from GCN into binarized collaborative filtering, which makes use of binary representation to improve the efficiency of online recommendation. However, binary codes are not only hard to be optimized but also likely to incur the loss of information during the training processing. Therefore, we propose a novel framework to convert the binary constrained optimization problem into an equivalent continuous optimization problem with a stochastic penalty. The binarized collaborative filtering model is then easily optimized by many popular solvers like SGD and Adam. The proposed algorithm is finally evaluated on three real-world datasets and shown the superiority to the competing baselines. 
	\end{abstract}
	
	\section{Introduction}
	Nowadays, recommender systerms are widely used in people's daily life~\cite{liu2011personalized,lian2016mutual,li2018learning}, but a growing scale of users and products renders recommendation challenging. Because implicit feedback is more common and easier to collect than explicit feedback, we concentrate on how to accelerate top-K recommendation based on implicit feedback. However, there are two challengings to address. Firstly,  
	compared with explicit feedback, implicit feedback is more difficult to utilize because of the lack of negative samples~\cite{pan2008one}. Secondly, generating top-k preferred items for each user is extremely time-consuming.
	
	For the first problem, recently, SpectralCF~\cite{zheng2018spectral} combined collaborative filtering model with graph convolutional network~\cite{henaff2015deep} to mine hidden interactions between users and items from spectral domain, which showed enormous potential for implicit feedback problem~\cite{zheng2018spectral}. However, SpectralCF ignores high-order feature interaction.
	
	For the second problem, for extracting top-K preferred items for each user, the time complexity of recommendation is $\mathcal{O}(MND+MN{\rm log}K)$ when there are $M$ users, $N$ items and $D$ dimensions in the latent space. Therefore, this is a critical efficiency bottleneck. However, it is necessary to timely update recommendation algorithms and the recommendation list because user interest evolves frequently. Fortunately, hash technique, encoding real-valued vectors/matrices into binary codes($e.g.,\{0,1\},\{-1,1\}$), is promising to address this challenge because inner product can be efficiently computed between binary codes via bit operation. Finding approximate top-K items can be even finished in sublinear or logarithmic time~\cite{wang2012semi,muja2009fast} by making use of index technique.
	
	Several methods applied hash techniques to recommendation. Some two-stage approximation methods like BCCF~\cite{zhou2012learning}, PPH~\cite{zhang2014preference}, CH~\cite{liu2014collaborative} incur large quantization loss~\cite{zhang2016discrete}, and a direct optimization model DCF~\cite{zhang2016discrete} is easy to fall into a local optimum because it is based on local search. To this end, to improve the accuracy of hashing-based recommender systems for implicit feedback, we propose a binarized collaborative filtering framework with distilling graph convolutional network. In the framework, we firstly train a CF-based GCN model~(GCN-CF) which can capture high-order feature interaction via cross operation. Following that, we distill the ranking information from the trained GCN-CF model into a binarized model~(DGCN-BinCF) with knowledge distillation technique~(KD~\cite{hinton2015distilling}). To be more specific, we introduce a novel distillation loss, which penalizes not only the discrepancy between distributions of positive items defined by GCN-CF and that defined by Bin-CF, but also the the discrepancy between distributions of sampled negative items. Noting that learning hash codes is generally NP-hard~\cite{haastad2001some}, approximation methods are appropriate choices but it may incur the loss of information during the training process. To this end, inspired by~\cite{dai2016binary}, we transform the binary optimization problem to an equivalent continuous optimization problem by imposing a stochastic penalty term. Therefore, any gradient-based solver can optimize the overall loss with ranking-based loss with knowledge distillation loss.
	
	Our contributions are summarized as follows:
	\begin{itemize}
		\item We propose a novel framework DGCN-BinCF to distill the ranking information from the proposed GCN-CF model into the binary model. To the best of our knowledge, DGCN-BinCF is the first model utilizing knowledge distilling to improve the performance of binarized model. We also improve GCN via adding a cross operation to aggregate users and items' own high-order features.
		\item We propose a generic method to relax the binary constraint problem to an equivalent bound-constrained continuous optimization problem. Hence, we can optimize the original problem by popular solvers directly.
		\item Through extensive experiments performed on three real-world datasets, we show the superiority of the proposed framework to the state-of-the-art baselines.
	\end{itemize}
	\section{Related Work}
	In this section, we review several works related to our task including GCN for recommender systems, recent hashing-based collaborative filtering methods and distilling knowledge techniques for ranking.
	\subsection{GCN for Recommender Systems}
	How to take advantage of the rich linkage information from the user-item bipartite graph is crucial for implicit feedback. Some work used GCN to solve it such as SpectralCF~\cite{zheng2018spectral}, GCMC~\cite{berg2017graph}, RMGCNN~\cite{monti2017geometric}, GCNWSRS~\cite{ying2018graph}, LGCN~\cite{gao2018large}, etc. (1)SpectralCF was the first model to learn from the Spectral domain of the user-item bipartite graph directly based on collaborative filtering. Because it could discover deep connections between users and items, it may alleviate cold-start problem. (2)GCMC combined GCN model with a graph auto-encoder to learn users' and items' latent factors. (3)RMGCNN proposed a matrix completion architecture combining a multi-graph convolutional neural network with a recurrent neural network. (4)GCNWSRS focused on how to apply GCN model for web-scale recommendation tasks effectively, like billion of items and hundreds of millions of users. (5)LGCN proposed a sub-graph training strategy to save memory and computational resource requirements greatly. Its experiments showed it was more efficient as compared to prior approaches.
	\subsection{Discrete Hashing for Collaborative Filtering}
	A pioneer work was to exploit Locality-Sensitive Hashing~\cite{datar2004locality} to generate binary codes for Google News readers according to their click history~\cite{das2007google}. Then \cite{karatzoglou2010collaborative} proposed a method mapping users and items' latent factors into Hamming space to obtain binary representation. Later, following this, some two stage methods~\cite{zhou2012learning,zhang2014preference} which relaxed binary constraints at first and then quantified binary codes. Nonetheless, \cite{zhang2016discrete} proposed that those two-stage methods suffered from large quantization loss. Therefore, DCF proposed a method which could optimize binary codes directly. However, DCF optimizes binary codes via searching neighborhoods with the distance one. So it is easy to fall into local optima.
	\subsection{Distilling Knowledge for Ranking}
	\cite{hinton2015distilling} was the first one that proposed method "Knowledge Distilling", which trained a complex neural network firstly and then transferred the complex model to a small model. The role of the complex model is similar to a teacher, and the role of the small model is similar to a student. Following this, DarkRank~\cite{chen2018darkrank} proposed a method combining deep metric learning and "Learning to rank" technique with KD to solve pedestrian re-identification, image retrieval
	and image clustering tasks. In addition, \cite{tang2018ranking} applied KD with point-wise ranking on recommendation task. Unfortunately, it did not focus on implicit feedback problem and how to transfer unobserved interaction information.
	\section{Definitions and Preliminaries}
	Throughout the paper, we denote vectors by boldfaced lowercase letters and matrices by boldfaced uppercase letters. All vectors are considered as column vectors. Next, we define the following definitions in this paper:
	\begin{myDef}
		(Bipartite Graph)A user-item bipartite graph with $M+N$ vertices and $E$ edges is defined as $\mathcal{G}=\{\mathcal{U},\mathcal{I},\mathcal{E}\}$, where $\mathcal{U}$ and $\mathcal{I}$ are two disjoint vertex sets of user and item, and $M=|\mathcal{U}|$, $N=|\mathcal{I}|$. For each edge $e\in\mathcal{E}$, it has the form that $e=(u,i)$, where $u\in\mathcal{U}$ and $i\in\mathcal{I}$, which shows that there exists an interaction between user $u$ and item $i$ in the training set.
	\end{myDef}
	\begin{myDef}
		(Laplacian Matrix)Given a bipartite graph with $M+N$ vertices and $E$ edges, the laplacian matrix $\textbf{\textit{L}}$ is defined as $\textbf{\textit{L}}=\textbf{\textit{D}}^{-1/2}\textbf{\textit{A}}\textbf{\textit{D}}^{-1/2}$, where $A$ is the adjacent matrix and $\textbf{\textit{D}}$ is the $(M+N)\times (M+N)$ diagonal degree matrix defined as $\textbf{\textit{D}}_{nn}=\sum_{j}\textbf{\textit{A}}_{nj}$.
	\end{myDef}
	Our work focuses on recommendation based on implicit feedback, where we only observe whether a user has viewed or clicked an item. We denote $\mathcal{I}_{i}^{+}$ as the set of all items clicked by user $i$ and denote $\mathcal{I}_{i}^{-}$ as the set of remaining items. 
	\subsection{Binary Collaborative Filtering}
	Matrix factorization maps users and items onto a joint $D$-dimensional latent space, where user embedding matrix is represented by $\textbf{\textit{P}}=[\textbf{\textit{p}}_{1},\cdots,\textbf{\textit{p}}_{M}]^{'}\in\mathbb{R}^{M\times D}$ and item embedding matrix is represented by $\textbf{\textit{Q}}=[\textbf{\textit{q}}_{1},\cdots,\textbf{\textit{q}}_{N}]^{'}\in\mathbb{R}^{N\times D}$. However, binary collaborative filtering~(Bin-CF) maps users and items onto a joint $D$-dimensional Hamming space. Denoting $\bm{\Phi}=[\bm{\phi}_{1},\cdots,\bm{\phi}_{M}]^{'}\in \{-1,1\}^{M\times D}$ and $\bm{\Psi}=[\bm{\psi}_{1},\cdots,\bm{\psi}_{M}]^{'}\in \{-1,1\}^{N\times D}$ as user and item's binary codes respectively, for implicit feedback, the Bin-CF problem is formulated as follows:
	\begin{align}
	\label{Bin-CF}
	\notag\mathop{\arg\min}_{\bm{\Phi},\bm{\Psi}}\notag\mathcal{L}_{Bin-CF} &=\sum_{(i,j,j^{'})\in \mathcal{D}}-\rm ln \sigma(\textbf{\textit{p}}_{\textit{i}}^{T}(\textbf{\textit{q}}_{\textit{j}}-\textbf{\textit{q}}_{\textit{j}^{'}}))
	\rm\\
	{\rm s.t.} \bm{\Phi}&=H(\textbf{\textit{P}}),\bm{\Psi}=H(\textbf{\textit{Q}})
	\end{align}
	where $H(\cdot)$ is a hash function:$\mathbb{R}\to\{-1,1\}$
	\subsection{Binary-Continuous Equivalent Transformation}
	Let us consider the following generic binary program firstly,
	\begin{align}
	\label{original binary}
	\notag \min &f(\textbf{\textit{x}})\\
	\rm s.t. \textbf{\textit{x}}\in&\{-1,1\}^{d}
	\end{align}
	and a transformed problem,
	\begin{align}
	\label{transformed binary}
	\notag \min f(\textbf{\textit{x}})&+\beta g(\textbf{\textit{x}})\\
	\rm s.t. \textbf{\textit{x}}\in&[-1,1]^{d}
	\end{align}
	where $g(\cdot)$:$\mathbb{R}^{d}\to\mathbb{R}$ is a penalty term for $f(\textbf{\textit{x}})$ and $\beta$ is its penalty coefficient. \cite{giannessi1998connections,lucidi2010exact} show that the above two problems are equivalent when certain conditions hold.
	
	\begin{lemma}
		\label{lemma1}
		Denote $||\cdot||$ be a chosen norm. Suppose the following conditions hold:
		\begin{enumerate}[1)]
			\item When $\textbf{\textit{x}}\in[-1,1]^{d}$, $f(\textbf{\textit{x}})$ is bounded. In addition, there exists an open set $A\supset\{-1,1\}^{d}$ and real positive number $\eta$, such that for $\forall\textbf{\textit{x}}_{1},\textbf{\textit{x}}_{2}\in A$, the following $H\ddot{o}lder$ condition is satisfied:
			\begin{equation}
			|f(\textbf{\textit{x}}_{1})-f(\textbf{\textit{x}}_{2})|\le\eta||\textbf{\textit{x}}_{1}-\textbf{\textit{x}}_{2}||
			\end{equation}
			\item $g(\cdot)$ satisfies:
			\begin{enumerate}[(a)]
				\item $g(\cdot)$ is continuous on $[-1,1]^{d}$
				\item $\forall \textbf{\textit{x}}\in \{-1,1\}^{d}, g(\textbf{\textit{x}})=0;\forall \textbf{\textit{x}}\in (-1,1)^{d}, g(\textbf{\textit{x}})>0$
				\item $\forall\textbf{\textit{y}}\in\{-1,1\}^{d}$, there exits a neighborhood $S(\textbf{\textit{y}})$ of $\textbf{\textit{y}}$ and a real positive number $\epsilon(\textbf{\textit{y}})$, such that:
				\begin{equation}
				\forall \textbf{\textit{x}}\in S(\textbf{\textit{y}})\cap (-1,1)^{d},g(\textbf{\textit{x}})\ge\epsilon(\textbf{\textit{y}})||\textbf{\textit{x}}-\textbf{\textit{y}}||
				\end{equation}
			\end{enumerate}
		\end{enumerate}
		Then there exits a real value $\eta_{0}$, such that $\forall \eta>\eta_{0}$, problem \ref{original binary} and problem \ref{transformed binary} are equivalent.
	\end{lemma}
	It can be verified that $g(\textbf{\textit{x}})=|||\textbf{\textit{x}}|-\textbf{1}||_{F}^{2}$ satisfies above conditions, and we adopt it as the penalty term.
	\section{Binarized Collaborative Filtering with Distilling Graph Convolutional Network}
	For binarized collaborative filtering for implicit feedback problem as shown in Eqn.\ref{Bin-CF}, there are three problems to solve. Firstly, the interaction information between users and items is extremely sparse. Secondly, a lot of information is lost during learning binary codes. Thirdly, binary optimization is general NP-hard, so we must adopt an efficient approximate method to solve it. We propose a novel framework-Binarized Collaborative Filtering with Distilling Graph Convolutional Network to deal with the aforementioned problems. Because GCN model can mine hidden connection information between users and items in user-item graph spectral domain, we train a GCN-based collaborative filtering model~(GCN-CF) to solve the first problem. Then we utilize knowledge distillation to transfer the ranking information from GCN-CF into the binary model to make up for information loss. Finally, we propose a method to transform the binary optimization to a continuous optimization problem to solve the binary optimization problem.     
	\subsection{GCN-based Collaborative Filtering}
	Following SpectralCF, our graph convolutional operation is shown as the following:
	\begin{align}
	\left[
	\begin{matrix}
	\textbf{\textit{U}}^{(k+1)}\\
	\textbf{\textit{V}}^{(k+1)}
	\end{matrix}
	\right]
	=\rho((\textbf{\textit{I}}_{M+N}+\textbf{\textit{L}})
	\left[
	\begin{matrix}
	\textbf{\textit{U}}^{(k)}\\
	\textbf{\textit{V}}^{(k)}
	\end{matrix}
	\right]\bm{\Theta}^{(k)})
	\end{align}
	where $\textbf{\textit{L}}$ is the Laplacian matrix of the bipartite graph $\mathcal{G}$. $\textbf{\textit{I}}_{M+N}$ is an identity matrix, $\rho(\cdot)$ is an activation function and $\bm{\Theta}^{(k)}$ is a layer-specific trainable filter parameter. The proposed convolution operation as shown in Eqn.~(6) is denoted as $\rm sp$$(\textbf{\textit{X}};\textbf{\textit{L}},\bm{\Theta})$.In this model, we set it as a two-layer GCN. According to Eqn.~(1), similarity to matrix factorization~(MF) methods, it does not take advantage of the user's own and the item's own high-order interaction, which limits the performance of GCN. Inspired by CrossNet~\cite{wang2017deep}, we define the \textbf{cross operation}($\rm cross\_op$) to fix the problem. The cross operation can be formulated as 
	\begin{align} \textbf{\textit{x}}_{k+1}=\textbf{\textit{x}}_{k}\textbf{\textit{w}}_{k}^{T}\textbf{\textit{x}}_{k}+\textbf{\textit{x}}_{k}
	\end{align}
	where $\textbf{\textit{x}}_{k}$ is a user or item's embedding vector and $\textbf{\textit{w}}_{k}$ is a parameter vector. The term $\textbf{\textit{x}}_{k}\textbf{\textit{w}}_{k}^{T}\textbf{\textit{x}}_{k}$ takes the place of the term $\textbf{\textit{x}}_{0}\textbf{\textit{w}}_{k}^{T}\textbf{\textit{x}}_{k}$ in CrossNet, which leads to obtaining higher-order interactions than CrossNet when setting the same iterations. In addition, the time complexity of the proposed cross operation is still the same as CrossNet's. The improved GCN model can be vectorized as follows:
	\begin{align}
	\label{FP1}
	\left[
	\begin{matrix}
	\textbf{\textit{U}}^{(1)}\\
	\textbf{\textit{V}}^{(1)}
	\end{matrix}
	\right]
	&=\rm sp
	(\left[
	\begin{matrix}
	\textbf{\textit{U}}^{(0)}\\
	\textbf{\textit{V}}^{(0)}
	\end{matrix}
	\right];\textbf{\textit{L}},\bm{\Theta}^{(0)})\\
	\label{FP2}
	\left[
	\begin{matrix}
	\textbf{\textit{U}}^{(2)}\\
	\textbf{\textit{V}}^{(2)}
	\end{matrix}
	\right]&=\rm diag ((\left[
	\begin{matrix}
	\textbf{\textit{U}}^{(1)}\\
	\textbf{\textit{V}}^{(1)}
	\end{matrix}
	\right]\circ\textbf{\textit{W}}_{1})\textbf{1})
	\left[
	\begin{matrix}
	\textbf{\textit{U}}^{(1)}\\
	\textbf{\textit{V}}^{(1)}
	\end{matrix}
	\right]+\left[
	\begin{matrix}
	\textbf{\textit{U}}^{(1)}\\
	\textbf{\textit{V}}^{(1)}
	\end{matrix}
	\right]\\
	\label{FP3}
	\left[
	\begin{matrix}
	\textbf{\textit{U}}^{(3)}\\
	\textbf{\textit{V}}^{(3)}
	\end{matrix}
	\right]&=\rm diag ((\left[
	\begin{matrix}
	\textbf{\textit{U}}^{(2)}\\
	\textbf{\textit{V}}^{(2)}
	\end{matrix}
	\right]\circ\textbf{\textit{W}}_{2})\textbf{1})
	\left[
	\begin{matrix}
	\textbf{\textit{U}}^{(2)}\\
	\textbf{\textit{V}}^{(2)}
	\end{matrix}
	\right]+\left[
	\begin{matrix}
	\textbf{\textit{U}}^{(2)}\\
	\textbf{\textit{V}}^{(2)}
	\end{matrix}
	\right]\\
	\label{FP4}
	\left[
	\begin{matrix}
	\textbf{\textit{U}}^{(4)}\\
	\textbf{\textit{V}}^{(4)}
	\end{matrix}
	\right]
	&=\rm sp
	(\left[
	\begin{matrix}
	\textbf{\textit{U}}^{(3)}\\
	\textbf{\textit{V}}^{(3)}
	\end{matrix}
	\right];\textbf{\textit{L}},\bm{\Theta}^{(1)})
	\end{align}
	where "$\circ$" represents Hadamard product, "\textbf{1}" is a column vector whose elements are all 1 and $\textbf{\textit{W}}_{1},\textbf{\textit{W}}_{2}\in\mathbb{R}^{(M+N)\times D}$ are weight matrices. Moreover, we add batch normalization~\cite{ioffe2015batch} before Eqn.\ref{FP1} and Eqn.\ref{FP4}.
	
	In order to make full use of features from every layer of GCN, we follow SpectralCF and concatenate them into the final latent factors of users and items as:
	\begin{align}
	\label{U}
	\textbf{\textit{U}}^{T}=[\textbf{\textit{U}}^{(0)},\textbf{\textit{U}}^{(1)},\textbf{\textit{U}}^{(4)}]\\
	\label{V}
	\textbf{\textit{V}}^{T}=[\textbf{\textit{V}}^{(0)},\textbf{\textit{V}}^{(1)},\textbf{\textit{V}}^{(4)}]
	\end{align}
	In terms of the loss function, we employ the popular and effective BPR loss~\cite{rendle2009bpr}. In particular, given a user matrix $\textbf{\textit{U}}$ and an item matrix $\textbf{\textit{V}}$ as shown in Eqn.\ref{U} and Eqn.\ref{V}, the loss function of GCN-CF is given as
	\begin{align}
	\label{GCN-CF}
	\notag\mathcal{L}_{GCN-CF}= &\sum_{(i,j,j^{'})\in \mathcal{D}}-\rm ln \sigma(\textbf{\textit{u}}_{\textit{i}}^{T}(\textbf{\textit{v}}_{\textit{j}}-\textbf{\textit{v}}_{\textit{j}^{'}}))\\
	&+\lambda(||\textbf{\textit{U}}||_{F}^{2}+||\textbf{\textit{V}}||_{F}^{2})
	\end{align}
	where $\textbf{\textit{u}}_{\textit{i}}$ and $\textbf{\textit{v}}_{\textit{j}}$ denote $i$th and $j$th rows of $\textbf{\textit{U}}$ and $\textbf{\textit{V}}$ respectively; $\lambda$ is the regularization coefficient. Negative sample $j^{'}$ is sampled from $\mathcal{I}_{i}^{-}$ randomly and the training data $D$ is generated as $\mathcal{D}=\{(i,j,j^{'})|i\in\mathcal{U}\land j\in\mathcal{I}_{i}^{+}\land j^{'}\in\mathcal{I}_{i}^{-}\}$.
	\subsubsection{Distilling GCN into Binarized Collaborative Filtering}
	In this model, we distill the ranking information in GCN-CF model and transfer it to a simple binary collaborative filtering model via a mixed objective function. The model is denoted by DGCN-BinCF for short. The \textit{key motivation} of the distillation in DGCN-BinCF is two-fold. On one hand, we consider the distribution of positive~(negative) samples in the binary model should be close to that in the GCN-CF. On the other hand, the differences between positive and negative samples should become far enough in DGCN-BinCF. However, because the BPR model can guarantee that positive samples have higher scores than negative samples', negative samples are assigned much lower probability than positive samples' if we consider the distribution of both positive and negative samples at the same time. Thus, we consider positive and negative samples in GCN-CF separately.
	
	Specifically, to distill the ranking information in GCN-CF, we hope the positive~(negative) items of one user have the approximately same order in binary model and continuous model. For instance, if user $u$'s preference for the three items $i,j,k$ is ranked as $[i,j,k]$ in GCN-CF model, we hope that the rank keeps $[i,j,k]$ in the binary model. According to ListNet~\cite{cao2007learning}, we can characterize sorting information in the following ways
	\begin{align}
	\label{rankloss}
	\notag&\mathcal{L}_{rank}=\sum_{(i,j)\in\mathcal{D}^{+}}-\frac{{\rm exp}(\frac{\textbf{\textit{u}}_\textit{i}^{T}\textbf{\textit{v}}_\textit{j}}{T})}{\sum_{j\in\mathcal{D}^{+}}{\rm exp}(\frac{\textbf{\textit{u}}_\textit{i}^{T}\textbf{\textit{v}}_\textit{j}}{T})}{\rm log}(\frac{{\rm exp}(\frac{\textbf{\textit{p}}_\textit{i}^{T}\textbf{\textit{q}}_\textit{j}}{T})}{\sum_{j\in\mathcal{D}^{+}}{\rm exp}(\frac{\textbf{\textit{p}}_\textit{i}^{T}\textbf{\textit{q}}_\textit{j}}{T})})\\
	&+\sum_{(i,j)\in\mathcal{D}^{-}}-\frac{{\rm exp}(\frac{\textbf{\textit{u}}_\textit{i}^{T}\textbf{\textit{v}}_\textit{j}}{T})}{\sum_{j\in\mathcal{D}^{-}}{\rm exp}(\frac{\textbf{\textit{u}}_\textit{i}^{T}\textbf{\textit{v}}_\textit{j}}{T})}{\rm log}(\frac{{\rm exp}(\frac{\textbf{\textit{p}}_\textit{i}^{T}\textbf{\textit{q}}_\textit{j}}{T})}{\sum_{j\in\mathcal{D}^{-}}{\rm exp}(\frac{\textbf{\textit{p}}_\textit{i}^{T}\textbf{\textit{q}}_\textit{j}}{T})})
	\end{align}
	\begin{algorithm}
		\KwIn{Bipartite graph:$\mathcal{G}$;dimension of latent factor in GCN-CF:$D$;dimension of latent factor in DGCN-BinCF:$d=3D$;Laplacian matrix:$\textbf{\textit{L}}$;temperature:$T,\tau$;coefficient:$\lambda,\alpha,\eta,\nu$}
		Initialize GCN-CF model;\\
		\Repeat
		{\text{Convergence}}
		{	Randomly sampling from unobserved items to generate training set $\mathcal{D}$;\\
			Forward propagation according to Eqn.\ref{FP1} $\thicksim$ \ref{FP4};\\
			Concatenate $\textbf{\textit{U}}$ and $\textbf{\textit{V}}$ via Eqn.\ref{U} and \ref{V};\\
			Update parameters of GCN-CF model by Adam optimization;
		}
		Initialize DGCN-BinCF model;\\
		\Repeat
		{\text{Convergence}}
		{	Randomly sampling from unobserved items to generate training set $\mathcal{D}$;\\
			Forward propagation according to Eqn.\ref{loss2};\\
			Update parameters of DGCN-BinCF model by Adam optimization;
		}
		\caption{DGCN-BinCF Algorithm}
		\label{alg:DGCN-BinCF}
	\end{algorithm}
	where $\mathcal{D}^{+}=\{(i,j)|i\in\mathcal{U}\land j\in\mathcal{I}_{i}^{+}\}$,  $\mathcal{D}^{-}=\{(i,j^{'})|i\in\mathcal{U}\land j^{'}\in\mathcal{I}_{i}^{-}\}$ and $T$ is the temperature parameter. In Eqn.\ref{rankloss}, we convert the items' score list to probability distributions via softmax function, and utilize cross entropy for penalizing the discrepancy. According to \cite{hinton2015distilling}, combining $\mathcal{L}_{Bin-CF}$ with $\mathcal{L}_{rank}$ as a multi-task learing problem can transfer the ranking knowledge to the binary model. It's worth mentioning that since the magnitudes of the gradients produced by the $\mathcal{L}_{rank}$ scale as $1/T^{2}$, it is necessary to multiply them by $T^{2}$ when mixing $\mathcal{L}_{rank}$ and $\mathcal{L}_{Bin-CF}$. So the loss function of DGCN-BinCF is formulated as 
	\begin{align}
	\label{DGCN-BinCF}
	\mathcal{L}_{DGCN-BinCF} = \mathcal{L}_{Bin-CF}+\alpha T^{2}\mathcal{L}_{rank}
	\end{align}
	$\mathcal{L}_{DGCN-BinCF}$ is denoted as $\mathcal{L}(\textbf{\textit{P}},\textbf{\textit{Q}};\textbf{\textit{U}},\textbf{\textit{V}}, \alpha, T)$ for short. $\textbf{\textit{P}}$ and $\textbf{\textit{Q}}$ are user and item embedding matrices in DGCN-BinCF respectively, and $\textbf{\textit{U}}$, $\textbf{\textit{V}}$ are trained user and item embedding matrices of GCN-CF. The loss $\mathcal{L}_{rank}$ encodes the first motivation and the loss $\mathcal{L}_{Bin-CF}$ encodes the second motivation.
	
	For the binary optimization problem, it is a direct method to use ${\rm tanh}(x/t)$ to approximate sign function, where $t$ is a small temperature. But \cite{li2018towards} points that setting a small temperature will harm the optimization process. \cite{courbariaux2015binaryconnect} mentions that generating binary codes stochastically is a finer and more correct averaging process than generating binary codes via sign function. Hence, we generate binary codes via sampling from the Bernoulli distribution. More specifically, given $\textbf{\textit{x}}\in (-1,1)^{d}$, its corresponding binary code is $\tilde{\textbf{\textit{x}}}=\textbf{\textit{x}}+\bm{\varepsilon}$, where $\bm{\varepsilon}$ is a random variable which only can be $\textbf{1}-\textbf{\textit{x}}$ and $-\textbf{1}-\textbf{\textit{x}}$, and $P(\bm{\varepsilon}_{i}=1-\textbf{\textit{x}}_{i})=\rm sigmoid(\textbf{\textit{x}}_{i}/\tau)$, $P(\bm{\varepsilon}_{i}=-1-\textbf{\textit{x}}_{i})=1-\rm sigmoid(\textbf{\textit{x}}_{i}/\tau)$. Here, $\tau$ is temperature parameter. To force the noise to be small, we add the expectation of noise as a penalty term. Therefore, the DGCN-BinCF can be transformed into the following optimization problem:
	\begin{align}
	\label{loss1}
	\notag\min \mathcal{L}&(\rm tanh(\textbf{\textit{P}})+\bm{\varepsilon}_{\textbf{\textit{P}}},\rm tanh(\textbf{\textit{Q}})+\bm{\varepsilon}_{\textbf{\textit{Q}}};\textbf{\textit{U}},\textbf{\textit{V}}, \alpha, T)\\&+\nu (\mathbb{E}(||\bm{\varepsilon}_{\textbf{\textit{P}}}||_{F}^{2})+\mathbb{E}(||\bm{\varepsilon}_{\textbf{\textit{Q}}}||_{F}^{2}))
	\end{align}
	where $(\bm{\varepsilon}_{\textbf{\textit{P}}})_{ij}\thicksim Bernoulli({\rm tanh}(\textbf{\textit{P}}_{ij}))$, $(\bm{\varepsilon}_{\textbf{\textit{Q}}})_{ij}\thicksim Bernoulli({\rm tanh}(\textbf{\textit{Q}}_{ij}))$. We use the $\rm \textbf{tanh}$ function to bound the value of $\textbf{\textit{P}},\textbf{\textit{Q}}$ between -1 and 1.
	
	According to Lemma \ref{lemma1}, Eqn \ref{loss1} can be rewritten as 
	\begin{align}
	\label{loss2}
	\notag\min &\mathcal{L}(\rm tanh(\textbf{\textit{P}}),\rm tanh(\textbf{\textit{Q}});\textbf{\textit{U}},\textbf{\textit{V}}, \alpha, T)\\
	\notag&+\nu(\mathbb{E}(||\bm{\varepsilon}_{\textbf{\textit{P}}}||_{F}^{2})+\mathbb{E}(||\bm{\varepsilon}_{\textbf{\textit{Q}}}||_{F}^{2}))\\
	&+\beta\Big(\sum_{i} g\big({\rm tanh}(\textbf{\textit{p}}_{i})\big) + 
	\sum_{j} g\big({\rm tanh}(\textbf{\textit{q}}_{j})\big)\Big)
	\end{align}
	Here we adopt $g(\textbf{\textit{p}}_{i})=|||\textbf{\textit{p}}_{i}|-\textbf{1}||_{2}^{2}$ which can be validated satisfying the conditions as the penalty term. Eqn.\ref{loss2} can be optimized by any gradient-based optimization methods directly. The whole training processing is summarized in Algorithm\ref{alg:DGCN-BinCF}.
	\section{Experiments}
	In this section, we evaluate our proposed DGCN-BinCF framework with the aim of answering the following research questions.
	\begin{enumerate}
		\item Does the recommendation performance of the proposed DGCN-BinCF framework outperforms the state-of-the-art hashing-based recommendation methods?
		\item Whether our proposed GCN-CF is effective? 
		\item Whether distilling ranking information helps learning binary model?
		\item Whether this proposed framework can converge well?
	\end{enumerate}
	
	We introduce the experimental settings firstly and then answer the above questions in following sections.
		
	\subsection{Experiment Settings}
	\subsubsection{Dataset}
	We use three public real datasets including \textit{MovieLens1M}, \textit{MovieLens10M} and \textit{Yelp} to evaluate the proposed algorithm.
	%
	%
	Because the three datasets are explicit feedback data, to convert them into implicit feedback data, we set all ratings as positive samples. In addition, due to the extreme sparsity of them, we then filter users who have less than 20 ratings and remove items that are rated by less than 20 users. Table \ref{tab:dataset} summaries the filtered datasets. For each user, we sampled randomly 50$\%$ positive samples as training and the remaining as test. We repeated five random splits and reported the averaged results.
	\begin{table}[h]
		\setlength\tabcolsep{4pt}
		\centering
		\begin{tabular}{ccccc}
			\toprule
			Dataset     & \#User & \#Item & \#Rating  & Density \\ \midrule
			{MovieLens1M} &  6,022   & 3,043  &  995,154  & 5.43\%  \\
			{MovieLens10M}  & 69,878 & 10,681  & 10,000,054 & 1.34\%  \\
			{Yelp}      & 9,235 & 7,353 &  423,354  & 0.62\%  \\ \bottomrule
		\end{tabular}
		\caption{Statistics of datasets}
		\label{tab:dataset}
	\end{table}
	\subsubsection{Comparison Methods}
	To evaluate the performance of DGCN-BinCF for hashing-based recommender systems, we compare DGCN-BinCF with 3 very popular and state-of-art methods:~DCF, BCCF and PPH. DCF solves the binary optimization problem directly via bit-wise optimization. BCCF and PPH are two-stage methods.
	
	To measure the effectiveness of the improved GCN model, we compare GCN-CF with SpectralCF. And we compare DGCN-BinCF with the binary model $\mathcal{L}_{Bin-CF}$. To show the role of KD loss in binary optimization, $\mathcal{L}_{Bin-CF}$ is optimized by our proposed relaxation method as well.
	
	\subsubsection{Evaluation Metric}
	To evaluate the recommendation system performance, we choose four widely used ranking-based metric: (1)~NDCG~(Normalized Discounted Cumulative Gain), (2)~Recall, , and~(3)~MAP~(Mean Average Precision). We predicted the top-K preferred items from test set for each user in our experiments.
	\subsubsection{Parameter Settings}
	In our experiments, we set the regularization coefficient $\lambda=0.001$ in GCN-CF in all dataset. For DGCN-BinCF, we set temperature $T=1$, $\tau=0.2$, the weight $\alpha=10$, and the penalty coefficients $\beta=0.001$, $\nu=0.001$ in the three datasets. In addition, we set the dimension of users and items' latent factor of GCN-CF 16 in \textit{MovieLens10M} and 64 in the other two datasets. The learned matrices $\textbf{\textit{U}}$ ,$\textbf{\textit{V}}$ in GCN-CF are used as the initialization of $\textbf{\textit{P}}$ ,$\textbf{\textit{Q}}$ in DGCN-BinCF. All parameters of SpectralCF are set according to \cite{zheng2018spectral}.
	
	Besides, for DCF, BCCF and PPH, we held-out evaluation means on splits of training data randomly to tune the optimal hyper-parmenters via grid search. $\alpha$ and $\beta$ in DCF are tuned among the set $\{1e-4,1e-3,\cdots,1e1\}$. $\lambda$ in BCCF is tuned among the set $\{0.01,0.03,\cdots,0.09\}$ and $\lambda$ in PPH is tuned among the set $\{0.01,0.5,1,2,4,8,16\}$.
	\begin{table}[h]
	\setlength\tabcolsep{3pt}
		\centering
		\begin{tabular}{lccc}
			\toprule
			& Recall@100 &  MAP@100& 
			NDCG@100\\
			\midrule
			DCF &  0.0416  & 0.0101  & 0.0558\\
			BCCF &  0.1234  & 0.0720 &  0.1724\\
			PPH &  0.0277 &  0.0027  & 0.0249\\
			DGCN-BinCF & 0.3059  & 0.1187  & 0.3061\\
			\bottomrule
		\end{tabular}
		\caption{Item Recommendation Results(MovieLens1M)}
		\label{tab:movielens1M}
	\end{table}
	\begin{table}[h]
	\setlength\tabcolsep{3pt}
		\centering
		\begin{tabular}{lccc}
			\toprule
			&  Recall@100 &  MAP@100& 
			NDCG@100\\
			\midrule
			DCF &  0.0791  & 0.0180 &   0.0809\\
			BCCF &  0.0789  & 0.0267  & 0.0869 \\
			PPH &  0.0978  & 0.0123  & 0.0695 \\
			DGCN-BinCF &  0.2405  & 0.0537  & 0.1895\\
			\bottomrule
		\end{tabular}
		\caption{Item Recommendation Results(MovieLens10M)}
		\label{tab:movielens10M}
	\end{table}
	\begin{table}[h]
	\setlength\tabcolsep{3pt}
		\centering
		\begin{tabular}{lccc}
			\toprule
			&Recall@100 &  MAP@100& 
			NDCG@100\\
			\midrule
			DCF &  0.0661  & 0.0053  & 0.0382\\
			BCCF &   0.0966 &  0.0122 &  0.0658 \\
			PPH & 0.0627 & 0.0051 & 0.0361 \\
			DGCN-BinCF &   0.1738  & 0.0192  & 0.1008\\
			\bottomrule
		\end{tabular}
		\caption{Item Recommendation Results(Yelp)}
		\label{tab:yelp}
	\end{table}
	\subsection{Comparison with Baselines}
	Although hashing-based recommendation has significant advantages of both time and storage, it often incurs low accuracy recommendation because binary codes have limited representation ability and lose a lot of information compared with real-valued recommender systems. DGCN-BinCF is to improve the accuracy of recommendation.
	
	In this part we will answer the first question. We compare the recommendation accuracy of DGCN-BinCF with three state-of-art binary recommendation methods including DCF, BCCF and PPH on the three datasets. Table \ref{tab:movielens1M}, Table \ref{tab:movielens10M} and Table \ref{tab:yelp} summary the results.
	
	The three tables show that DGCN-BinCF has much better performance than all baselines on the three datasets. This is because we train the improved GCN model GCN-CF firstly to discover the deep interactions between users and items and then transfer the ranking information to a binary model, DGCN-BinCF loses less information than baseline models. In addition, the binary optimization problem is optimized directly by the proposed penalty terms, which leads to less quantization loss. Therefore, DGCN-BinCF has great advantages over DCF, BCCF and PPH.
	\subsection{The Effectiveness of GCN-CF}
	It is mentioned that SpectralCF did not consider aggregating users and items' own high-order feature, which may limit its representation ability. In this part, we will answer the second question.
	
\begin{figure}[htbp]
		\centering
		\includegraphics[width=0.8\columnwidth]{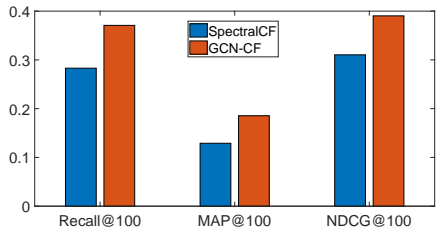}
		\caption{The performance of SpectralCF and GCN-CF}
		\label{fig:SCF_GCN-CF}
	\end{figure}

	We implement the experiment in MovieLens1M dataset. We utilize three metrics to evaluate the performance of SpectralCF and GCN-CF respectively. Figure \ref{fig:SCF_GCN-CF}
	shows the results. In two histograms, the orange column is the performance of GCN-CF and the blue one represents the results of SpectralCF. From the histograms, it is clear to observe that GCN-CF has great improvement~(over 20\%) for every metric compared with SpectralCF, which shows the effectiveness of GCN-CF.
	\subsection{The Effectiveness of Distillation}
	Because lots of useful information loses during learning the binary representation, it is vital to utilize the ranking information from the trained GCN-CF model as supplements for learning discrete codes. In this part, we investigate the role of ranking information for binary optimization. To implement the experiment, we consider optimizing Eqn.\ref{Bin-CF} directly by adding the proposed penalty term. In the other word, we set $\alpha=0$ in $\mathcal{L}_{DGCN-BinCF}$, and compare its results with DGCN-BinCF. We test the two methods in the MovieLens1M dataset and evaluate them via the four ranking metrics.
	
\begin{figure}[htbp]
		\centering
		\includegraphics[width=0.8\columnwidth]{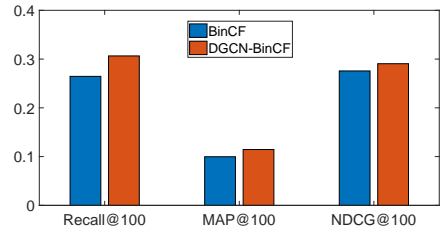}
		\caption{The comparison between BinCF and DGCN-BinCF}
		\label{fig:BinCF_DGCN-BinCF}
	\end{figure}

	Figure \ref{fig:BinCF_DGCN-BinCF} reports the comparison results. The blue bar represents BinCF and the orange bar means DGCN-BinCF model. The two histograms show that for all evaluation indicators, DGCN-BinCF outperforms BinCF by 10\%. Thus we conclude that the distillation method helps the model learn high quality binary representation.
	\begin{figure}[htbp]
		\centering
		\subfigure{
			\includegraphics[width=0.47\columnwidth]{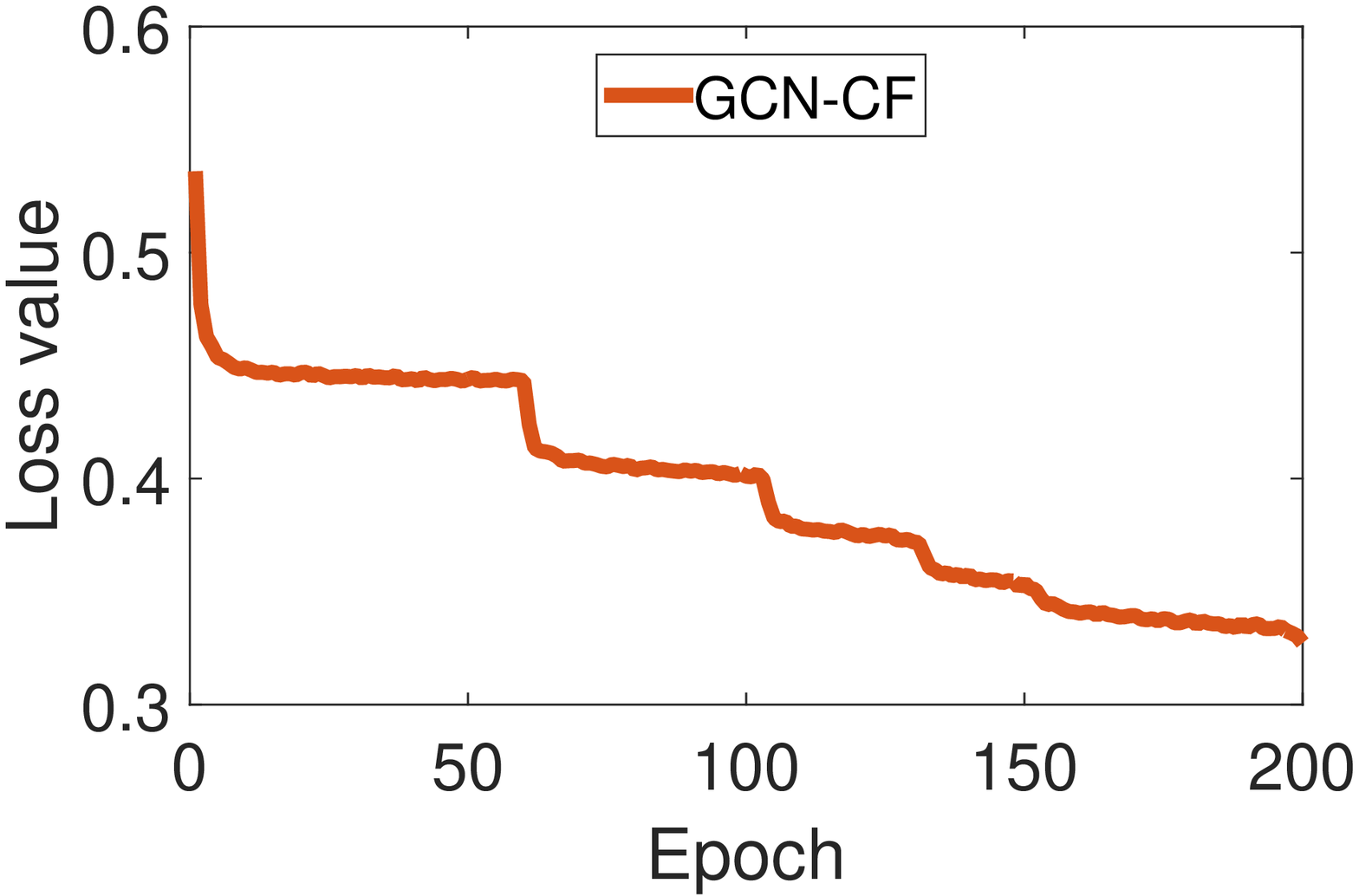}
		}
		\subfigure{
			\includegraphics[width=0.47\columnwidth]{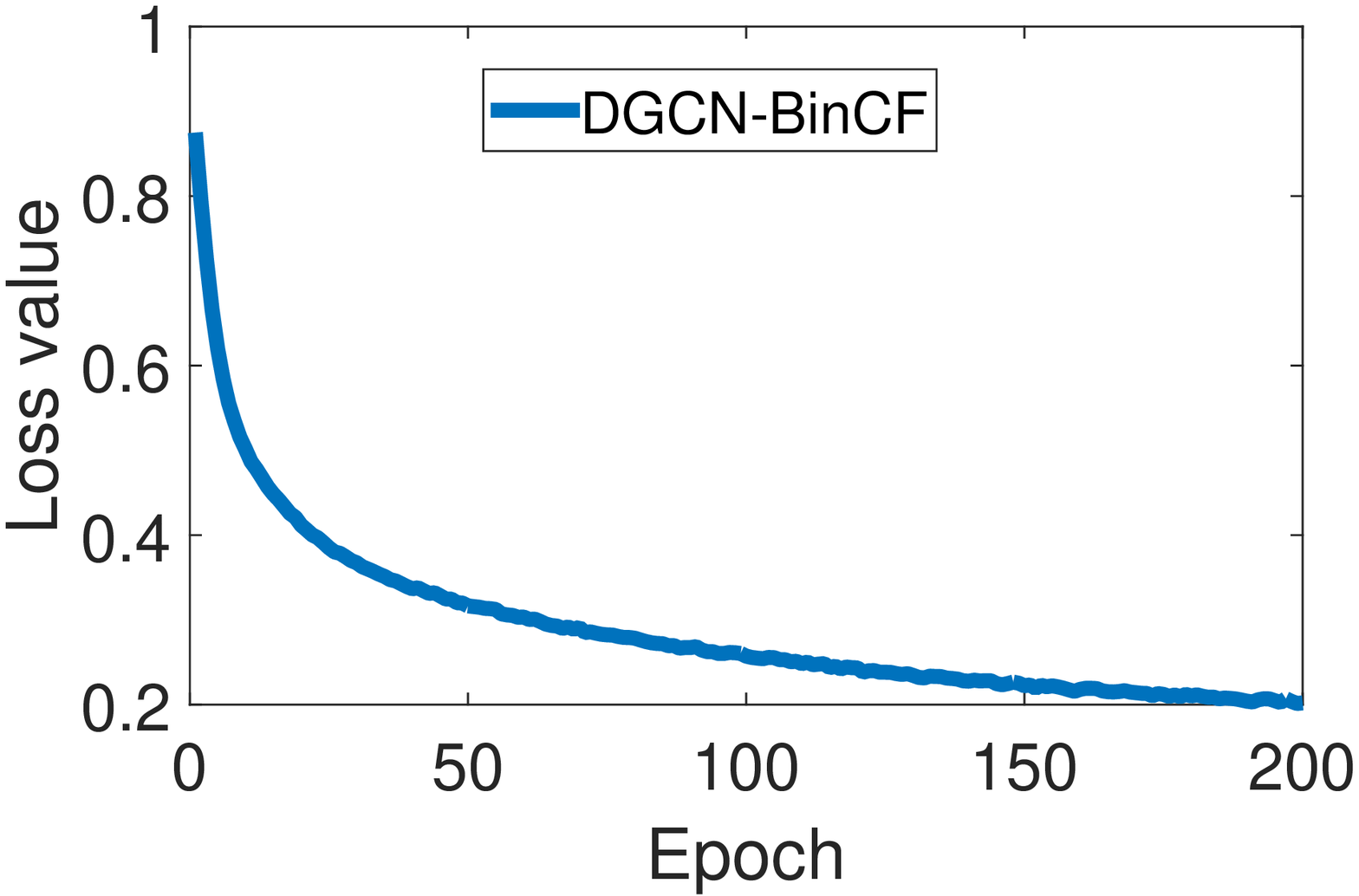}
		}
		\caption{The left one is the loss-epoch figure of GCN-CF model;the right one is the loss-epoch figure of DGCN-BinCF model.}
		\label{fig:loss-epoch}
	\end{figure}
	\subsection{Convergence}
	In this section, we will answer the forth question. Because deep models and discrete optimization may diverge, we test the convergence of GCN-CF and DGCN-BinCF model.
	
	To test the convergence of our proposed model GCN-CF and DGCN-BinCF, we implement the experiment on \textit{MovieLens1M}. We record the value of Eqn.\ref{GCN-CF} and Eqn.\ref{loss2} with the change of epoch respectively. In this experiment, we set the maximum number of iterations 200. Figure \ref{fig:loss-epoch} shows the convergence of two models. It is observed that loss value of GCN-CF decreases and the DGCN-BinCF converges greatly. 

	\section{Conclusion}
	In this paper, we propose a hash-based method DGCN-BinCF to accelerate implicit feedback recommendation. Because implicit feedback lacks negative samples and learning binary codes loses a lot of information, we train the model GCN-CF, which aggregates users' and items' own high-order feature, to mine rich connection information, and then distill the ranking information from GCN-CF into the binary model. In addition, we propose a method utilizing penalty terms to learning binary codes based on gradient descent directly. The experiments on three real-world datasets show the great superiority of our framework.
	\section*{Acknowledgements}
	The work was supported in part by grants from the National
	Natural Science Foundation of China (Grant No. U1605251,
	61832017, 61631005 and 61502077).
	{
			\small
		\bibliographystyle{named}
		\bibliography{ijcai19}
	}
\end{document}